\documentclass[apjl]{emulateapj}

\usepackage{mathptmx}
\usepackage{txfonts}
\usepackage[T1]{fontenc}
\usepackage{ae,aecompl}
\usepackage{booktabs}
\usepackage[flushleft]{threeparttable}

\usepackage{graphicx}
\usepackage{wasysym}
\usepackage{hyperref}

\newif\ifdumbeddown
\dumbeddowntrue

\newif\ifpaper
\paperfalse

\hypersetup{
pdftitle={GW170104 and INTEGRAL},
pdfsubject={GW170104 and INTEGRAL},
pdfauthor={Savchenko et al},
pdfkeywords={GW170104, INTEGRAL}
}

\newcommand{\ec}[1] {{#1}~erg~cm$^{-2}$}

\usepackage{color}

\newcommand{\software}[1] {\textit{#1}}

\newcommand{\updated}[1] {#1}
\newcommand{\updatedsecond}[1] {#1}

\begin{document}

\shorttitle{INTEGRAL observations of GW170104}
\shortauthors{V. Savchenko et al.}

\title{INTEGRAL observations of GW170104}

\author{ V.~Savchenko$^{1}$, 
C.~Ferrigno$^{1}$, 
E.~Bozzo$^{1}$, 
A.~Bazzano$^{2}$, 
S.~Brandt$^{3}$, 
J.~Chenevez$^{3}$, 
T.~J.-L.~Courvoisier$^{1}$, 
R.~Diehl$^{4}$, 
L.~Hanlon$^{5}$, 
A.~von~Kienlin$^{4}$, 
E.~Kuulkers$^{6}$, 
P.~Laurent$^{7,8}$, 
F.~Lebrun$^{8}$, 
A.~Lutovinov$^{9,10}$, 
A.~Martin-Carillo$^{5}$, 
S.~Mereghetti$^{11}$, 
J.~P.~Roques$^{12}$, 
R.~Sunyaev$^{9,13}$, 
 and P.~Ubertini$^{2}$
\\
  }
\affil{  $^{1}$ISDC, Department of astronomy, University of Geneva, chemin d'\'Ecogia, 16 CH-1290 Versoix, Switzerland \\
 $^{2}$INAF-Institute for Space Astrophysics and Planetology, Via Fosso del Cavaliere 100, 00133-Rome, Italy \\
 $^{3}$DTU Space - National Space Institute Elektrovej - Building 327 DK-2800 Kongens Lyngby Denmark \\
 $^{4}$Max-Planck-Institut f\"{u}r Extraterrestrische Physik, Garching, Germany \\
 $^{5}$Space Science Group, School of Physics, University College Dublin, Belfield, Dublin 4, Ireland \\
 $^{6}$European Space Research and Technology Centre (ESA/ESTEC), Keplerlaan 1, 2201 AZ Noordwijk, The Netherlands \\
 $^{7}$APC, AstroParticule et Cosmologie, Universit\'e Paris Diderot, CNRS/IN2P3, CEA/Irfu, Observatoire de Paris Sorbonne Paris Cit\'e,\\
\hspace{0.05cm}10 rue Alice Domont et L\'eonie Duquet, 75205 Paris Cedex 13, France. \\
 $^{8}$DSM/Irfu/Service d'Astrophysique, Bat. 709 Orme des Merisiers CEA Saclay, 91191 Gif-sur-Yvette Cedex, France \\
 $^{9}$Space Research Institute of Russian Academy of Sciences, Profsoyuznaya 84/32, 117997 Moscow, Russia  \\
 $^{10}$Moscow Institute of Physics and Technology, Institutskiy per. 9, Dolgoprudny, Moscow Region, 141700, Russia\\
 $^{11}$INAF, IASF-Milano, via E.Bassini 15, I-20133 Milano, Italy \\
 $^{12}$Universit\'e Toulouse; UPS-OMP; CNRS; IRAP; 9 Av. Roche, BP 44346, F-31028 Toulouse, France \\
 $^{13}$Max Planck Institute for Astrophysics, Karl-Schwarzschild-Str. 1, Garching b. Munchen D-85741, Germany   }

\date{Accepted XXX. Received YYY; in original form ZZZ}

\label{firstpage}

\begin{abstract}
We used data from the INTErnational Gamma-Ray Astrophysics Laboratory
(INTEGRAL) to set upper-limits on the $\gamma$-ray and hard X-ray
prompt emission associated with the gravitational wave event
GW170104, discovered by the LIGO/Virgo collaboration.  The
unique omni-directional viewing capability of the instruments on-board
INTEGRAL allowed us to examine the full 90\% confidence level
localization region of the LIGO trigger. Depending on the particular
spectral model assumed and the specific position within this region,
the upper limits inferred from the INTEGRAL observations range from
$F_{\gamma}$=1.9$\times$10$^{-7}$~erg~cm$^{-2}$
to
$F_{\gamma}$=10$^{-6}$~erg~cm$^{-2}$
(75~keV~-~2~MeV energy range). This translates into a ratio between
the prompt energy released in $\gamma$-rays along the direction to the
observer and the gravitational wave energy of
E$_\gamma/$E$_\mathrm{GW}<$2.6$\times$10$^{-5}$.
Using the INTEGRAL results, we can not confirm the $\gamma$-ray
proposed counterpart to GW170104 by the AGILE team with the
MCAL instrument. The reported flux of the AGILE/MCAL event,
E2, is not compatible with the INTEGRAL
upper limits within most of the 90\% LIGO localization region.  There
is only a relatively limited portion of the sky where the sensitivity
of the INTEGRAL istruments was not optimal and the lowest allowed
fluence estimated for E2 would still be
compatible with the INTEGRAL results.  This region was also observed
independently by Fermi/GBM and AstroSAT, from which, as far as we are
aware, there are no reports of any significant detection of a
prompt high-energy event.
\end{abstract}


\maketitle



\section{Introduction}

The LIGO/Virgo collaboration reported a third significant
gravitational-wave (GW) event, GW170104, discovered on
2017-01-04 10:11:58.6 UTC. The false alarm probability
associated with the detection was less than one event over 70\,000
years \citep{LVC_GW170104_paper}. The LIGO 90\% confidence
localization region of GW170104 consisted of two elongated
arcs, each spanning over
120~deg.  The event was
associated with the merging of two black holes with masses of
31\small$^{+8.4}_{-6}$\normalsize~M$_\odot$ and
19\small$^{+5.3}_{-5.9}$\normalsize~M$_\odot$ at a distance of
880\small$^{+450}_{-390}$\normalsize~Mpc. GW170104 is thus
the most remote confirmed GW event discovered so far.

Following the announcement by the LIGO team, extensive follow-up
observations were carried out by a large number of facilities to
search for an electromagnetic counterpart. Results obtained from
on-going serendipitous observations were promptly reported as
well. The two telescopes on-board the Fermi satellite could not detect
any significant excess over the background that was spatially and
temporally compatible with the GW event
\citep{burns2017_gcn20365,fermi_gw170104}.  Fermi-GBM provided sky
coverage of
69.5\% at
the time of GW170104, enclosing
82.4\% of the LIGO
localization region. The upper limit derived from the Fermi-GBM
observations corresponds to a 1-second fluence spanning from
\ec{5.2$\times$10$^{-7}$} to
\ec{9.4$\times$10$^{-7}$} (in the 8-1000 keV
energy range and assuming a typical Band spectrum of a short
$\gamma$-ray burst, GRB).  A tighter upper limit on the fluence of the
event was reported by AstroSAT in a more restricted region of the sky
\citep{Bhalerao2017}. A non-detection at 95\% confidence level (c.l.)
was also reported by Konus-Wind \citep{Svinkin2017_gcn21158}.

One of the instruments on-board the AGILE satellite revealed an excess
over the instrument background (AGILE-GW170104)
that was roughly coincident in time with the GW event.  The estimated
signal to noise ratio (SNR) of the detection is
4.4 and the corresponding
\updatedsecond{post-trial coincidence probability is between
  2.4~$\sigma$ and
  2.7~$\sigma$}
\citep{Verrecchia2017}.

In this letter, we make use of the available data collected by the
instruments on-board INTEGRAL \citep{integral} to search for possible
hard X-ray and $\gamma$-ray counterparts to GW170104. We
summarize the most relevant capabilities of the INTEGRAL instruments
for these kinds of searches in Sect.~\ref{sec:instruments} and
describe all the obtained results in Sect.~\ref{sec:results}. We
discuss the non-detection of a counterpart to the GW event in the
INTEGRAL data with respect to the findings reported by the AGILE team
in Section~\ref{sec:agile}. Our conclusions are reported in
Section~\ref{sec:conclusions}.

\section{The INTEGRAL instruments and the follow-up of GW events}
\label{sec:instruments}


As extensively described by \citet{Savchenko2017a}, INTEGRAL provides
unique instantaneous coverage of the entire high-energy sky by
taking advantage of the synergy between its four all-sky detectors:
IBIS/ISGRI, IBIS/PICsIT, IBIS/Veto, and SPI-ACS. These provide
complementary capabilities for the detection of transient events
characterized by different durations, locations on the sky, and
spectral energy distributions.  In the case of the first GW event,
GW150914, the most stringent upper limit on the non-detection of an
electromagnetic counterpart in 75~keV to 2~MeV energy range with
INTEGRAL was obtained with the SPI-ACS \citep{Savchenko2016}, while the
peculiar localization of LVT151012 \citep{ligoo1} and its orientation
with respect to the INTEGRAL satellite required the combination of the
results from all detectors (together with a careful analysis of each
instrument's response and background) to achieve an optimized upper
limit. As we discuss in Sect.~\ref{sec:results}, it is again the
SPI-ACS that provides the most stringent upper limit on the high
energy emission from the non-detected electromagnetic counterpart to
GW170104.

The SPI-ACS \citep{spiacs} is made of 91 BGO (Bismuth Germanate,
Bi$_4$Ge$_3$O$_{12}$) scintillator crystals and it is the
anti-coincidence shield of the SPI instrument \citep{spi}. Besides its
main function of shielding the SPI germanium detectors, the ACS is
also used as a nearly onmidirectional detector of transient events,
providing a large effective area at energies above $\sim$75~keV. The
ACS data consist of event rates integrated over all the scintillator
crystals with a time resolution of 50 ms. No spectral and/or
directional information of the recorded events is available. The
typical number of counts per 50 ms time bin ranges typically from about 3000 to
6000. SPI-ACS
features a high duty cycle of $\sim$85\%\footnote{The reduction of 15\% is
  due to the fact that the INTEGRAL instruments are switched-off near
  the perigee of every satellite revolution. The INTEGRAL orbit was as
  long as three sidereal days until January 2015, but was later
  shortened to 2.7 to allow for a safe satellite disposal in 2029.}
and comprises events from the nearly complete high energy sky.

SPI is partially surrounded by the satellite structure and by the
other INTEGRAL instruments, which shield the incoming photons and thus also 
affect the response of the ACS in different directions. For this reason, 
the computation of the ACS response requires detailed
simulations which take into account the entire satellite structure.  We
developed a \software{GEANT3} Monte-Carlo model based on the INTEGRAL
mass model \citep{sturner03} and simulated the propagation of
monochromatic parallel beams of photons in the 50 keV-100 MeV energy 
range. For each energy, we simulated 3072 sky positions (16-side
\software{HEALPix}\footnote{http://healpix.sourceforge.net}
grid). This allows us to generate an instrumental response function
for any position in the sky, which can then be used to compute the expected
number of counts for a given source spectral energy distribution. As shown   
in our previous paper \citep{Savchenko2017a}, this response produces results 
for the bursts detected simultaneously by the SPI-ACS and other detectors 
(Fermi/GBM and Konus-Wind) that are consistent to an accuracy 
better than 20\%.

\section{INTEGRAL observations of GW170104}
\label{sec:results}

\begin{figure*}
\centering 
\vspace{0.5cm}
 \includegraphics[width=0.9\columnwidth, angle=0]{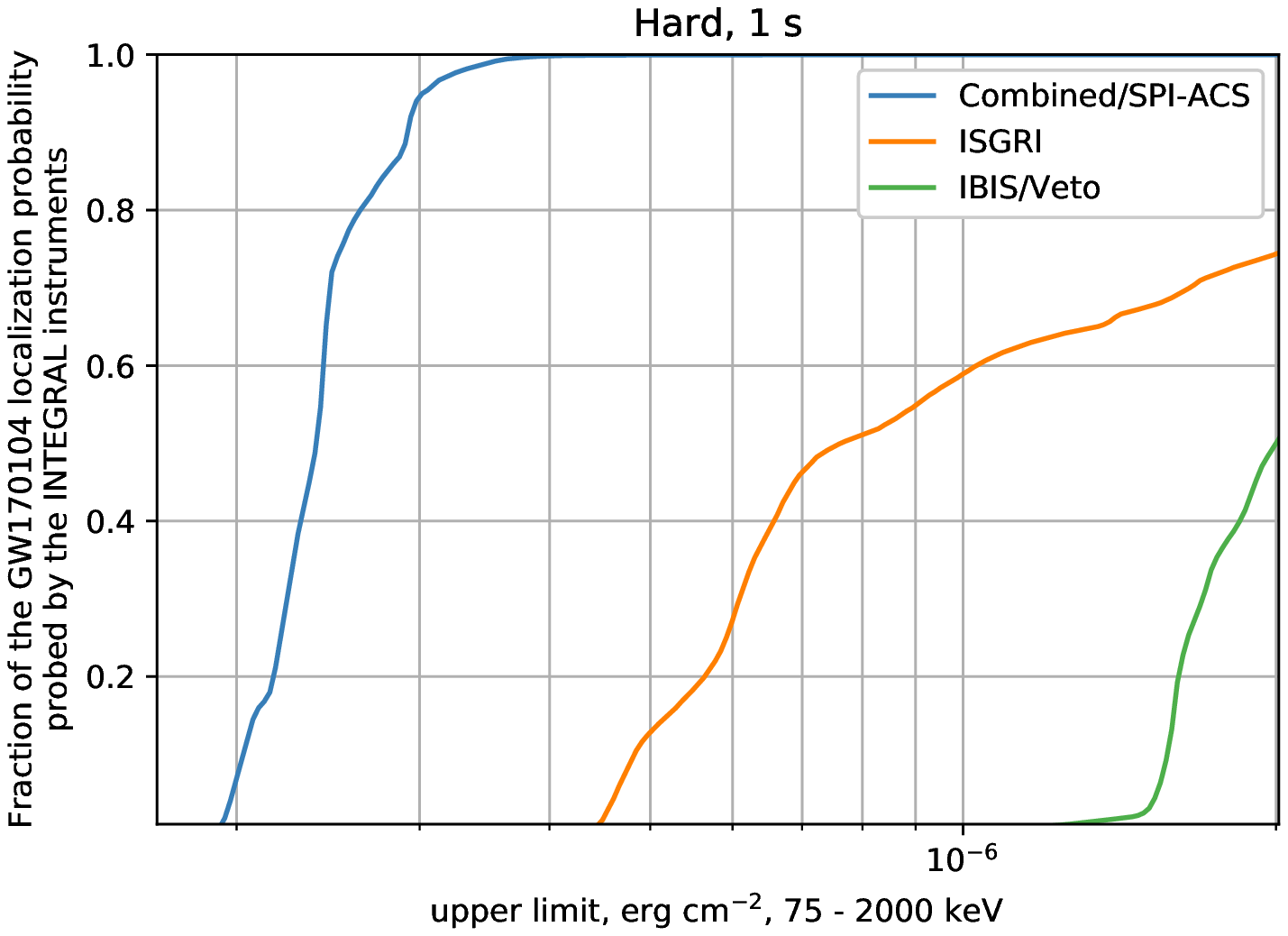} 
     \includegraphics[width=0.9\columnwidth, angle=0]{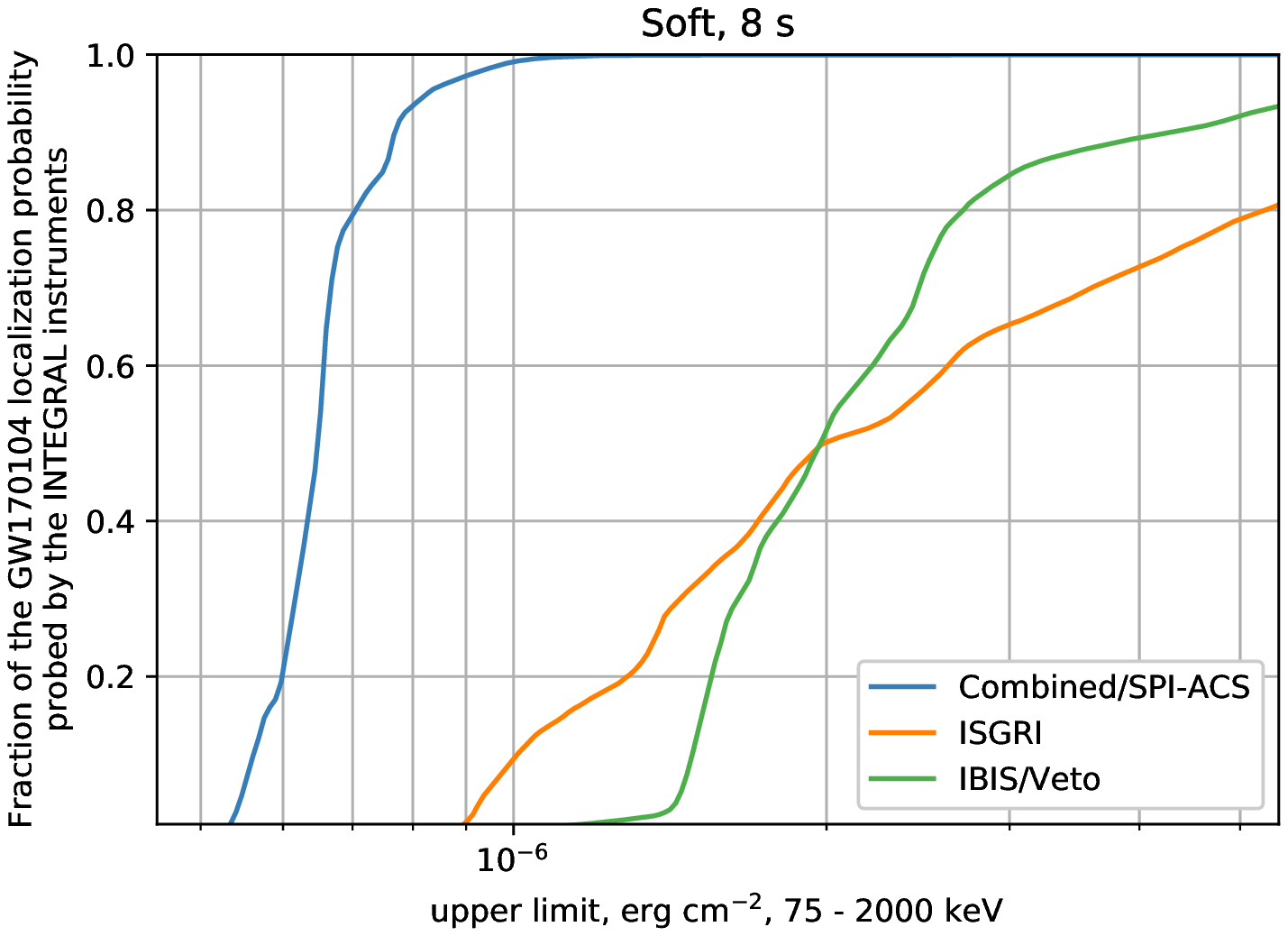} 
\caption{Plot of the fraction of the LIGO localization probability of
  GW170104 probed by the data of the different INTEGRAL
  instruments as a function of the upper-limit (3$\sigma$ c.l.) on the
  non-detected electromagnetic counterpart to the GW event. The figure
  on the left is for the case of the short-hard burst, while the
  figure on the right shows the case of a long-soft burst (see text
  for details). The "Combined/SPI-ACS" text in the label indicates
  that the results do not quantitatively change if only the SPI-ACS
  data are used to draw the blue solid line or if the independent
  contributions from the other instruments are also merged.}
\label{fig:sens_containment}
\end{figure*}

At the time of the GW170104
(2017-01-04 10:11:58.6 UTC, hereafter T$_0$) INTEGRAL
was fully operational and executing the pointing
ID.~176700040010 in the direction
of Cas A / Tycho SNR, far from the likely
localization region of the LIGO trigger. All instruments were
performing nominally, yielding a virtually constant and stable
background count rate from at least
T$_0$ - 2500 to
T$_0$ + 2500 ks. \updated{The SPI-ACS
  background count rate was about
  1.14$\times$10$^{5}$~counts~s$^{-1}$,
  which is higher than that observed during LVT151012 or GW150914 and
  close to the maximum value ever observed in SPI-ACS data during the
  INTEGRAL mission lifetime (excluding the time intervals affected by
  Solar flares). There are two reasons for the high background
  recorded at the time of GW170104: the 11-years Solar
  activity cycle, which is close to its minimum, and the day-scale
  variations of the instrumental background, which have been commonly
  observed since the early stages of the instrument operations. The
  enhanced background rate decreases the sensitivity of INTEGRAL
  instruments by as much as 30\%, when compared to the most favorable
  conditions and much less, when compared to our reports on LVT151012
  and GW150914. However, it should be noticed that the effects of
  background fluctuations on the sensitivity are typically smaller
  than those due to sky location.} At the time of GW170104, the Earth
was relatively distant from INTEGRAL, casting a small shadow of
49.0~deg$^2$ on
the instrument field-of-view (equivalent to
0.12\% of
the sky) and occulting only about
0.032\% of
LIGO event localization probability.  In the remaining part of this
region, the SPI-ACS sensitivity was close to optimal. Thus, this
instrument allowed us to carry out the most accurate search for any
electromagnetic counterpart to GW170104. For a fraction of
the 90\% LIGO localization region, the IBIS sensitivity, including
both ISGRI and PICsIT, \citep{theibis} approached that of the SPI-ACS,
but we checked that adding these data did not significantly improve
our results.  Therefore we do not extensively comment on the IBIS data
but report for completeness in Fig.~\ref{fig:sens_containment} a
comparison between the contributions provided by the SPI-ACS,
IBIS/Veto, and ISGRI in searching for an electromagnetic counterpart
of GW170104. In this figure, we estimated for each value of
the upper limit the integrated fraction of the entire LIGO
localization region of the GW event that is probed by the data of the
different INTEGRAL instruments.  The SPI-ACS is clearly able to
provide the deepest limits in the entire portion of the sky where the
LIGO localization probability is significantly larger than zero.

The INTEGRAL Burst Alert System (IBAS) \citep{mereghetti03} routinely
inspects the INTEGRAL SPI-ACS and IBIS/ISGRI lightcurves in real time,
searching for significant deviations from the background and producing
automatic triggers.  The closest IBAS trigger to GW170104
occurred on 2017-01-04 22:12:40 (T$_0$ +
43241~s)
and was classified as a cosmic ray event, thus unlikely to be related
to the LIGO trigger.

The closest event identified as a possible GRB in INTEGRAL data
occured at 2017-01-05 06:14:06 with a SNR of
9.3 and a duration of
5~seconds. The astrophysical
nature of this event was confirmed by simultaneous observations of
Konus-Wind \citep{Svinkin2017_gcn20406}, AstroSAT
\citep{Sharma2017_gcn20389}, POLAR \citep{Marcinkowski2017_gcn20387},
and a combined IPN analysis \citep{Svinkin2017_gcn20406}. This was
classified as a regular long GRB
(GRB170105) with an optical afterglow
that could also be independently found in the ATLAS follow-up
observations of GW170104
\citep[ATLAS17aeu;][]{Tonry2017_gcn20377,Melandri2017_gcn20735,stalder2017,Bhalerao2017}.
INTEGRAL observations contributed to the triangulation which allowed
the establishing the association between
GRB170105 and ATLAS17aeu
\citep{Svinkin2017_gcn20406}. In general, INTEGRAL data are
particularly useful to retrospectively search for GRB events, owing to
its competitive and consistent omnidirectional sensitivity, stable
background, and high duty cycle \citep[see e.g. a recent case studied
by][]{Whitesides2017}.  GRB170105 was
later found to be likely unrelated to GW170104
\citep{stalder2017,Bhalerao2017}.

\begin{figure}
  \centering 
  \vspace{0.5cm} \includegraphics[width=0.99\columnwidth]{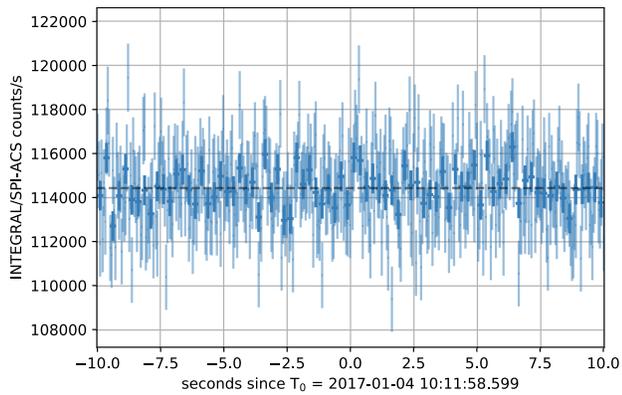}
  \caption{Zoom of the INTEGRAL/SPI-ACS lightcurve in the $\pm$10~s
    time interval around the LIGO detection of GW170104.
    Light blue symbols represent the measurements at the natural
    instrument time resolution of 50 ms, while dark blue points
    correspond to the data rebinned at a resolution of 250 ms. The
    dashed black curve represents the average instrument background
    obtained from a much longer span of data.}
  \label{fig:lightcurve}
\end{figure}

We also inspected the SPI-ACS and IBIS light curves, focusing on
a time interval of $\pm$500 s around T$_0$ and probing 5 different
time scales in the range 0.05-100 s. The latter were selected to be
representative of the dynamical time scale of the accretion occuring
in a coalescing compact binary \citep[e.g.][]{lee07}.  We did not
find any obvious detection of a significant signal temporally
coincident with the GW event.  A zoom of the SPI-ACS lightcurve around
the time of the LIGO trigger is shown in Figure~\ref{fig:lightcurve}.

Following the approach in \cite{Savchenko2016,Savchenko2017a} and the
non-detection of any significant electromagnetic counterpart to
GW170104 in the INTEGRAL data, we derived the corresponding
upper limits assuming the cases of (i) a \textit{short-hard} burst,
i.e.  a 1~s-long event characterized by a cut-off power-law spectral
energy distribution with parameters
$\alpha=-0.5$,
$E_{peak}=600~$keV; (ii) a
\textit{long-soft} burst, i.e.  an 8-s long event whose spectral
energy distribution is described by the Band model \citep{band93} with
parameters $\alpha=-1$,
$\beta=-2.5$, and
$E_{peak}=300~$keV. \updated{While
  the reference short GRB duration of 1~s is close to the peak of the
  short GRB duration distribution, the 8~s time scale for the
  long-soft GRB is motivated by the sampling rate of IBIS/Veto, in analogy with the approach 
  presented previously by \cite{Savchenko2017a}.}

\begin{figure}
  \centering 
  \vspace{0.5cm}
   \includegraphics[width=0.99\columnwidth]{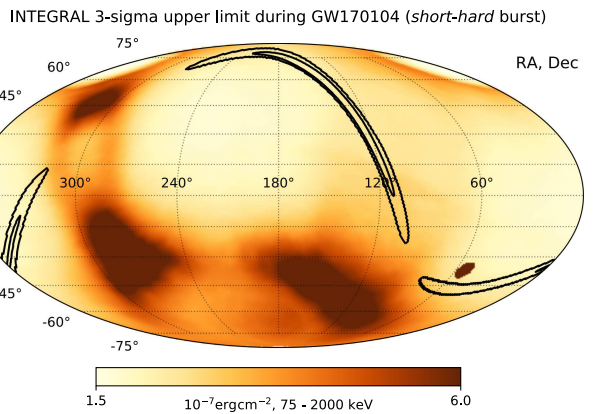}
  \caption{Estimated 3$\sigma$ upper limits on the 75-2000~keV flux of the non detected 
  electromagnetic counterpart to GW170104 as derived from the SPI-ACS data assuming the case of a 
  short-hard burst. The black contours show the most accurate localization of the GW event 
  at 50\% and 90\% c.l., as provided by the LALInference \citep{LVC_GW170104_paper}.}
  \label{fig:sky_hard}
\end{figure}

\begin{figure}
  \centering 
  \vspace{0.5cm}
   \includegraphics[width=0.99\columnwidth]{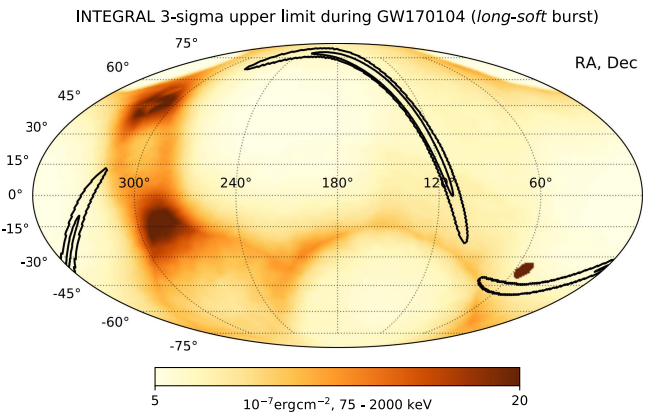}
  \caption{Same as Fig.~\ref{fig:sky_hard} but in the case of a long-soft burst.} 
  \label{fig:sky_soft}
\end{figure}

\updated{To calculate the 3-$\sigma$ upper limits, we fold the spectral
  model through the instrument response for each sky location and
  adjust the model normalization until the predicted number of counts
  is equal to three times the standard deviation of the background
  counts in the considered time interval. The upper limit derived in
  this way corresponds also to the 3-$\sigma$ detection threshold, which
  is the generally recommended approach to compute upper limits
  corresponding to the non detection of astrophysical events
  \citep{Kashyap2010}. Our method complies to the commonly accepted
  upper limit definitions, used for example by the Fermi/GBM team
  \citep{fermi_gw170104}.}  The results obtained in these two cases
are shown in Fig.~\ref{fig:sky_hard} and \ref{fig:sky_soft}. The
estimated upper limits (75~keV~-~2~MeV) within the LIGO 90\%
localization region range from
$F_{\gamma}=$1.9$\times$10$^{-7}$~erg~cm$^{-2}$
to
3.5$\times$10$^{-7}$~erg~cm$^{-2}$
for a 1-second short hard GRB and from
$F_{\gamma}=$5.2$\times$10$^{-7}$~erg~cm$^{-2}$
to
10$^{-6}$~erg~cm$^{-2}$
for an 8-second event characterized by a typical long GRB spectrum.


Assuming the reference distance to the event of
D=880~Mpc \citep{LVC_GW170104_paper}, we can
derive an upper limit on the isotropic equivalent \updated{total
  energy released in the 75~-~2000~keV energy band in one second as}
$E_{\gamma} <
3.2\times10^{49}
\mathrm{erg}
\left(\frac{F_{\gamma}}{3.5\times10^{-7}
    \mathrm{erg~cm}^{-2}}\right)
\left(\frac{D}{880
    \,\mathrm{Mpc}}\right)^{2}$. The energy emitted in gravitational
waves can be estimated as $E_\mathrm{GW}$ =
3.6\small$^{+1.1}_{-1.3}$\normalsize$ \times 10^{54}$~erg.  The SPI-ACS upper
limits we reported above can constrain the fraction of energy emitted
in hard X-rays and $\gamma$-rays towards the observer during the GW
event to be
$f_{\gamma}~<~9\times10^{-6}$
in the case of the \textit{short-hard} burst, and
$f_{\gamma}~<~2.6\times10^{-5}$
in the case of the \textit{long-soft} one \updated{(in the
  75~-~2000~keV energy range)}.

\updated{While the limit on the fraction of the gamma-ray energy
  emitted in the energy range covered by SPI-ACS has the advantage of
 depending the least on the assumed source spectrum, it is of a general
  interest to estimate a limit on the bolometric luminosity.  In the
  1~-~10\,000~keV energy range that is conventionally used
  \citep[e.g.][]{Rowlinson2014,Pescalli2016}, we can constrain the
  total released electromagnetic energy and its ratio to the GW energy
  as
  $E_{1-10^{5}keV}~<~3.5\times10^{49}$
  ($f_{1-10^{5}keV}~<~9.8\times10^{-6}$)
  in the case of the \textit{short-hard} burst, and
  $E_{1-10^{5}keV}~<~1.3\times10^{50}$
  ($f_{1-10^{5}keV}~<~3.7\times10^{-5}$)
  in the case of the \textit{long-soft} one.}

\subsection{On the possible AGILE detection of an electromagnetic counterpart to GW170104} \label{sec:agile}

AGILE is an X-ray and $\gamma$-ray astronomical satellite of The
Italian Space Agency, launched in 2007. AGILE's scientific payload
comprises a pair-conversion telescope, capable of detecting photons in
the 30~MeV~-~100~GeV energy range (GRID), and a hard X-ray monitor
sensitive in the 18~-~60~keV energy range (SuperAGILE or
SA). Additionally, AGILE is able to observe bright impulsive events
from a large fraction of the unocculted sky with its mini-Calorimeter
(MCAL), operating in the energy band 0.4-100 MeV \citep{Tavani2008}.

\cite{Verrecchia2017} reported on observations carried out with the
MCAL at the time of GW170104.  These observations covered
only a fraction of the LIGO localization, due to the occultation of
the AGILE FoV caused by the Earth.  Several weak bursts were
identified in the AGILE/MCAL data around the time of
GW170104. Among them, the 32~ms-long burst
E2 was identified as a possible
$\gamma$-ray counterpart of the GW event. The reported trigger time is
at 0.46~$\pm$~0.05~s before T$_0$.

\begin{figure}
  \centering 
  \vspace{0.5cm}
  \includegraphics[width=\columnwidth, angle =0]{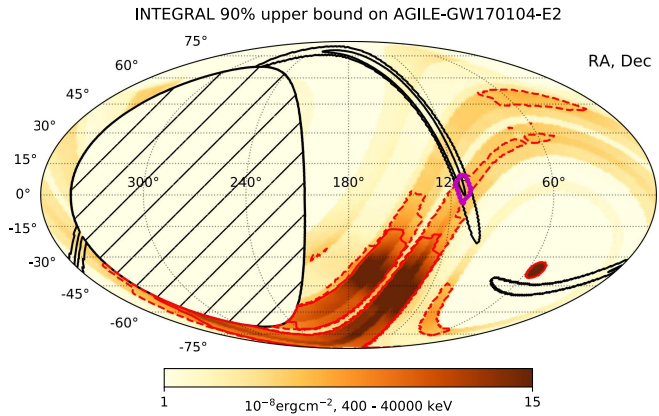} 
  \caption{Plot of the estimated lowest detectable fluence at 90\%
    c.l.  by the SPI-ACS for a 32~ms long burst going off at the time
    of AGILE-E2 in different positions of
    the sky (a spectral energy distribution with a slope of -2 has
    been assumed). \updated{The large dashed circle corresponds to the
      location occulted for AGILE by the Earth. The small dark circle represents the
      region occulted by the Earth to INTEGRAL.}  Solid red lines enclose
    the regions where the lowest detectable SPI-ACS fluence is higher
    than the best fit one
    (\ec{8.9$\times$10$^{-8}$}) obtained
    for the tentative AGILE counterpart of GW170104
    (i.e., the event E2).  Dashed red lines
    are used for the same comparison with the lower boundary of the AGILE
    fluence
    (\ec{5.9$\times$10$^{-8}$}).
    The thick magenta lines encircle the position of the sky within
    the 90\% LIGO localization region of GW170104 in which
    the minimum fluence reported for AGILE-GW170104 is
    compatible with the INTEGRAL results. }
  \label{fig:agilemap}
\end{figure}

\updated{Following the report by \citep{Verrecchia2017}, we
  investigated the INTEGRAL data to check for any confirmation of this
  detection.  We note that, unlike the upper limit presented in the
  previous Section (Figs~\ref{fig:sky_hard} and \ref{fig:sky_soft}),
  we need to compute the upper bound in the flux of any possible
  celestial event corresponding to the measured signal in SPI-ACS at
  the exact time of the AGILE putative event.  However, the INTEGRAL
  orbit is very elongated resulting in a sizable difference in a
  celestial signal arrival time, which depends on the unknown source
  sky location, reaching up to
  $\pm0.32$~s.  First,
  we computed for each position in the sky the time at which the event
  AGILE-GW170104 should have been observed by
  INTEGRAL. For each position in the sky at the proper trigger time, we show
  with a colour map in Fig.~\ref{fig:agilemap} the corresponding 90\%
  c.l.  values of the upper bound on the
  400~--~40000~keV
  fluence consistent with 
  the SPI-ACS count rate\footnote{Note that the 90\% c.l. was preferred to the
    3$\sigma$ approach to compare more easily the INTEGRAL and AGILE
    findings.}. The reported values are calculated assuming a
  32~ms-long event characterized by a power-law shaped spectral energy
  distribution with a slope of -2 \citep[as done
  in][]{Verrecchia2017}.  As SPI-ACS observes positive and negative
  count rate fluctuations in the background, all positions
  corresponding to a certain time delay between the INTEGRAL and AGILE
  locations define circularly-shaped regions in the sky within which
  the upper bound on the event flux is constant. This is the reason
  why the map of the upper bound values in Fig.~\ref{fig:agilemap}
  comprises stripes of different colors.  The source positions in the
  sky coincident with the direction toward AGILE as seen from INTEGRAL
  and the diametrically opposite direction correspond to the maximum
  absolute time delays.  Since the altitude of AGILE's orbit is much
  smaller than that of INTEGRAL's orbit, the direction from INTEGRAL
  towards AGILE is very close to the direction from INTEGRAL to Earth,
  and the circularly-shaped regions are all approximately centered on
  the position of the Earth (a small dark circle in
  Fig.~\ref{fig:agilemap}).  The median value of the fluence in sky
  locations compatible to the time delay between the spacecrafts} is
\ec{1.7$\times$10$^{-8}$} and
it does not exceed
\ec{7.1$\times$10$^{-8}$} in any
sky position enclosed within the LIGO 90\% localization region of
GW170104.  In Fig.~\ref{fig:agilemap}, we highlighted with
red contours the portions of the sky where the minimum detectable
fluence by INTEGRAL is consistent with the best fit (solid) and the
lowest allowed (dashed) fluence of
AGILE-GW170104 inferred from the AGILE data.

We found that there are no sky positions within the 90\% LIGO
localization region for which the best fit fluence of the AGILE event
is compatible with the INTEGRAL results.  There are, however,
positions within the 90\% LIGO localization region for which the
lowest allowed value of the fluence of
AGILE-GW170104 would still be compatible with
the INTEGRAL results (thick magenta contour on
Fig.~\ref{fig:agilemap}). The ensemble of these positions covers about
4.2\% of the
LIGO localization region and extends for a total of
77.5~deg$^2$. Note that
a few small regions enclosed within red dashed lines are sparsely
present in the color map of Fig.~\ref{fig:agilemap}. These are
positions in the sky for which the AGILE trigger time of
AGILE-GW170104 corresponds to positive count
rate fluctuations in the SPI-ACS lightcurve. We inspected each of
these fluctuations, but none of them exceeded a S/N of 1.5.

Taking together all these results, we cannot exclude that the event
AGILE-GW170104 is associated with the GW trigger
if it originated from a restricted number of positions in the sky
within the 90\% LIGO localization region. However, this detection is
compatible with the INTEGRAL results only if a fluence that is a
factor of
1.2
lower than the best fit value obtained from the AGILE data is
considered.

We noticed that the limited positions in the sky within the 90\% LIGO
localization region for which the AGILE/MCAL detection is compatible
with the INTEGRAL results were also accessible to the Fermi/GBM
\citep{fermi_gw170104, burns2017_gcn20365} and, in an even more
limited way, by the AstroSAT/CZTI \citep{Bhalerao2017}.  Further
analysis of the observations performed by these two facilities could
help to confirm or not the AGILE detection.

The conclusions above depend significantly on the assumed spectral
energy distribution of the event. A detailed description of the
spectral parameters of AGILE-GW170104 is not
provided by \cite{Verrecchia2017}, and thus we followed their
assumption of a power-law shaped energy distribution with a slope of
-2.  At the same time, the authors also indicated that
AGILE-GW170104 features similar timing and
spectral properties to the precursor of GRB090510. This weak precursor
was detected by both AGILE/MCAL and Fermi/GBM. It was also detected by
INTEGRAL/SPI-ACS with a S/N of 6.1,
even though the location in the sky was not covered with the optimal
sensitivity of the SPI-ACS.  By analyzing the response of this
instrument in the direction of GRB090510 and using the results
obtained from the observation of the precursor of the GRB, we were
able to derive a nearly model-independent conclusion that a similar
event occurring anywhere within the LIGO 90\% localization region of
GW170104, excluding the area invisible to AGILE, should
have been detected by the SPI-ACS with a median S/N of
13.0,
and certainly no lower than
4.6.

Finally, we stress that it is entirely possible that the AGILE/MCAL
event was a real weak short GRB going off in a region of the sky
covered with a low SPI-ACS sensitivity and completely unrelated to
GW170104 (i.e. outside the 90\% LIGO localization region).
Combining the area of the sky with unfavorable orientations for the
SPI-ACS observations and not occulted by the Earth for AGILE, we
inferred a remaining allowed region spanning about
3533~deg$^2$.

\section{Conclusions}
\label{sec:conclusions}

All GW events reported so far by LIGO were found to be most likely
associated with binary back hole mergers. The extensive
multi-wavelength follow-up campaigns carried out after each of these
discoveries led to the detection of at least two possible
electromagnetic counterparts to the GW events
\citep{gbmpaper,Greiner2016,Verrecchia2017}. Although none of these associations
was firmly confirmed, they led to discussion of exotic scenarios in
explaining EM emission in these mergers \citep[e.g.][]{perna16,loeb16,
  Woosley2016a, lyutikov2016}.  The INTEGRAL efforts to follow-up as
much as possible all relevant LIGO triggers will eventually help to
revealing which, if any, of these scenarios is applicable.  So far,
the INTEGRAL results have provided the most stringent upper limits on
any associated prompt hard X-ray and $\gamma$-ray emission in the
75~keV to 2~MeV energy range for each of the announced GW events when
INTEGRAL observations were available, challenging the possible
association of GW 150914 and GW 170104 with the tentatively reported
electromagnetic counterparts.

\section*{Acknowledgements}
Based on observations with INTEGRAL, an ESA project with instruments
and science data centre funded by ESA member states (especially the PI
countries: Denmark, France, Germany, Italy, Switzerland, Spain), and
with the participation of Russia and the USA. The SPI-ACS detector
system has been provided by MPE Garching/Germany. We acknowledge the
German INTEGRAL support through DLR grant 50 OG 1101. The Italian
INTEGRAL/IBIS team acknowledges the support of ASI/INAF agreement
n. 2013-025-R.0. AL and RS acknowledge the support from the Russian
Science Foundation (grant 14-22-00271). Some of the results in this
paper have been derived using the \software{HEALPix} \citep{healpix}
package. We are grateful the Fran\c cois Arago Centre at APC for
providing computing resources, and VirtualData from LABEX P2IO for
enabling access to the StratusLab academic cloud. Finally, we thank
the anonymous referee for the insightful comments.

\bibliographystyle{apj}

\label{lastpage}
\end{document}